\documentclass{JINST}

\let\ifpdf\relax
\usepackage{url}
\usepackage{tabularx}
\usepackage{multicol}
\usepackage{float} 

\newcommand{\figref}[1]{\figurename~\ref{#1}}
\newcommand{\secref}[1]{Section~\ref{#1}}

\newcommand{\tabref}[1]{Table~\ref{#1}}
\usepackage{booktabs}
\usepackage{xspace}
\usepackage{pbox}
\usepackage{mathrsfs}
\usepackage{amsmath}
\usepackage{verbatim}

\newcommand{\Er}{\ensuremath{\epsilon_{r}}\xspace}
\newcommand{\three}{\ensuremath{3\times1\times1 \mbox{m}^3}\xspace}

\title{First test of a high voltage feedthrough for liquid Argon TPCs connected to a 300 kV power supply}

\author{C.~Cantini$^a$, A.~Gendotti$^a$, L.~Molina Bueno $^a$, S.~Murphy$^a$,
   B. Radics$^a$, C.~Regenfus$^a$, Y-A.Rigaut$^a$, A.~Rubbia$^a$\thanks{Corresponding
    author.}, F.~Sergiampietri$^a$, T.~Viant$^a$ and S.~Wu$^a$~\\
  \llap{$^a$}ETH Zurich, Institute for Particle Physics,\\
  CH-8093 Z\"{u}rich, Switzerland\\
  E-mail: \email{Andre.Rubbia@cern.ch}}

\abstract{Voltages above a hundred kilo-volt will be required to generate the drift field of future very large liquid Argon Time Projection Chambers. The most delicate component is the feedthrough whose role is to safely deliver the very high voltage to the cathode through the thick insulating walls of the cryostat without compromising the purity of the argon inside. This requires a feedthrough that is typically meters long and carefully designed to be vacuum tight and have small heat input. Furthermore, all materials should be carefully chosen to allow operation in cryogenic conditions. In addition, electric fields in liquid argon should be kept below a threshold to reduce risks of discharges.
The combination of all above requirements represents significant challenges from the design and manufacturing perspective. In this paper, we report on the successful operation of a feedthrough satisfying all the above requirements. The details of the feedthrough design and its manufacturing steps are provided. 
Very high voltages up to unprecedented voltages of -300 kV could be applied during long periods repeatedly. A source of instability was observed, which was specific to the setup configuration which was used for the test and not due to the feedthrough itself. 
}


\begin{document}

\section{Introduction}\label{sec:introduction}
It is well established that liquid Argon Time
Projection Chambers (LAr TPCs) are the best suited detectors to
address the next major questions in neutrino physics.  Future giant
liquid Argon TPCs, at the 10 kton level, are now at the design and
prototyping stage in the context of the Deep Underground Neutrino
Experiment (DUNE) \cite{Acciarri:2016crz}.  The key concept of the LAr TPCs is that they provide a complete 3D image of the interaction final
state particles over a wide range of energy, allowing for efficient
background rejection and good energy reconstruction. From experience with liquid argon TPCs, a drift field between 250 and 500 V/cm is required to efficiently drift the electrons through the liquid argon \cite{Amoruso:2004dy}. The DUNE far detectors and their prototypes, aiming at a maximum drift distance of several meters require bias voltages in the hundreds of kV
range at their cathodes. Our goal of reaching -300 kV is motivated by the innovative dual phase LAr TPC design which proposes a fully homogeneous detector with one single vertical drift volume. This approach offers the advantage of a detector free from any material inside the active volume. The key feature of the dual phase (liquid-vapour) operation is the amplification of the signal by means of charge avalanche in the vapour \cite{Badertscher:2013wm,Cantini:2013yba,Cantini:2014xza}. The amplification yields a larger
signal to noise ratio and an overall better image quality thereby allowing the possibility of constructing detectors with longer drift
distances which would otherwise be plagued by ionisation cloud diffusion and charge attenuation by traces of electronegative impurities in the liquid. Intensive R\&D is currently underway within the WA105 project~\cite{WA105_TDR} to construct and demonstrate the feasibility of large
scale dual phase TPCs. Parameters of five LAr TPCs presently considered within DUNE are summarised in \tabref{tab:DUNE_HV}: the \three dual phase \cite{SPSC-SR-2016}, the protoDUNE single phase \cite{NP04_prop}, the
protoDUNE dual phase\cite{WA105_TDR}, and the single and dual
DUNE 10 kt far detectors\cite{Acciarri:2016crz}.

\begin{table}[htb]
\renewcommand{\arraystretch}{1.2}
\begin{center}
\begin{tabular}{p{3cm}p{1.95cm}p{1.95cm}p{1.95cm}p{1.85cm}p{2cm}}
  \toprule
  Detector & active LAr volume [m$^3$] & number of drift regions & drift length [m] &\multicolumn{2}{l}{\parbox{4.5cm}{high voltage at cathode [kV] }}\\
  \cmidrule{5-6}
          & & & & 250 V/cm & 500 V/cm\\
  \midrule
   \three dual phase \cite{SPSC-SR-2016} & 23 & 1 & 1 & -25  & -50 \\
  protoDUNE single phase\cite{NP04_prop} &  216 & 2 & 3.6 & -90  & -180 \\ 
  protoDUNE dual phase\cite{WA105_TDR} &  216 & 1 & 6 & -150  & -300 \\
  DUNE 10 kt single phase\cite{Acciarri:2016crz}&  8640 & 4 & 3.6 & -90  & -180 \\
  DUNE 10 kt dual phase\cite{Acciarri:2016crz} &  8640 & 1 &12& -300 & -600  \\
  \bottomrule
       \end{tabular}
     \end{center}
     \caption{\label{tab:DUNE_HV} Parameters of the liquid argon TPCs in DUNE detectors. One drift region is defined
       as a free space between a cathode and the readout electrodes.}
   \end{table}
As can be seen a first milestone of -300 kV at the cathode would allow to drift 250 V/cm in the DUNE dual phase far detector and 500 V/cm in its prototype. At such high potentials the detector components must be carefully designed to minimise the local electric fields in liquid Argon. The dielectric rigidity of liquid Argon has a wide range of values depending on its purity and cryogenic conditions. 
A reference value of 1.4 MV/cm for the dielectric strength of liquid argon was first measured and reported in \cite{Swan1}. Since then, 
it has however been observed that breakdown in liquid can occur at lower electric field due to the formation of bubbles (see \cite{Ushakov:2007}). In a recent test, we showed that a sustained stable electric field of -100 kV over one centimeter in liquid Argon was achievable at equilibrium \cite{Bay:2014jwa}. However, when the liquid was boiling breakdowns occurred as low as 40 kV/cm, supporting the idea that absence of bubbles is critical to ensure safe operation. On the other hand, in order to increase stability with respect to bubble formation which can never be totally excluded in large systems, electric fields in liquid argon should be kept below the threshold of 40~kV/cm. Similar results and conclusions  with a different electrode configuration were reported in \cite{Blatter:2014wua}.

\begin{figure}[htb]
     \centering
    \includegraphics[width=0.7\textwidth,clip=true]{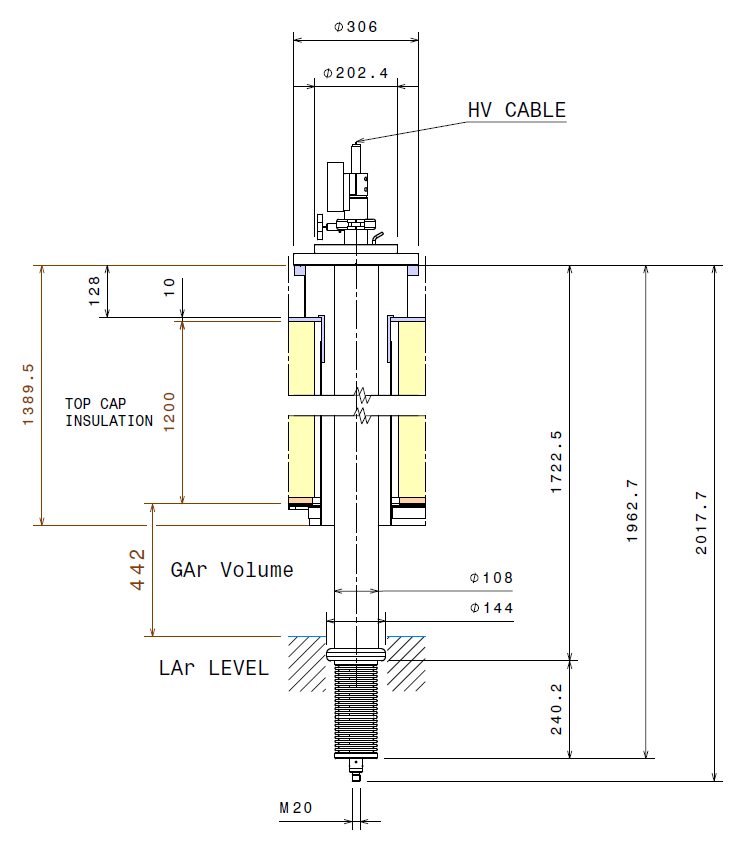}
     \caption{Drawing of the 300 kV high voltage feedthrough inserted through the cryostat hosting the WA105 dual phase \three LAr-TPC at CERN. All dimensions are given in mm.}
     \label{fig:HVFTin311}
   \end{figure}

   In order to polarise the cathode at the required high voltage three
   main components are necessary. The power supply, the external cable
   and the feedthrough whose role is to transport the high
   voltage from the cable through the cryostat insulation until below the
   liquid Argon level where the electrical connection to the cathode
   can be made. Up to -300 kV we can rely on commercial power supplies and cables which are described in the next paragraph. A feedthrough, capable of delivering such high voltage to the detector, however needs to be custom designed. This 300 kV feedthrough should have sufficient length to penetrate through the thick passive insulation of the cryostat and cross the Argon vapour
   before reaching the liquid. We show in \figref{fig:HVFTin311} a drawing of the 300 kV feedthrough inserted in the cryostat hosting a dual phase \three LAr-TPC currently in the final stage of construction at CERN. 
   In this configuration, the feedthrough crosses 1.2 meters of insulation and 44 cm of Argon vapour before being immersed in the liquid. The distances are of the same order for the protoDUNE dual and single phase TPCs which means that the detectors require a feedthrough with a length $\geq 2$ m. 
   
\section{Generation and transport of very high voltage to the
     liquid Argon} \label{sec:GenTransportHV} 
\subsection{The high voltage feedthrough} \label{sec:HVFT}
\subsubsection{Design}

\begin{figure}[htb]
  \centering
  \includegraphics[width=1\textwidth,clip=true]{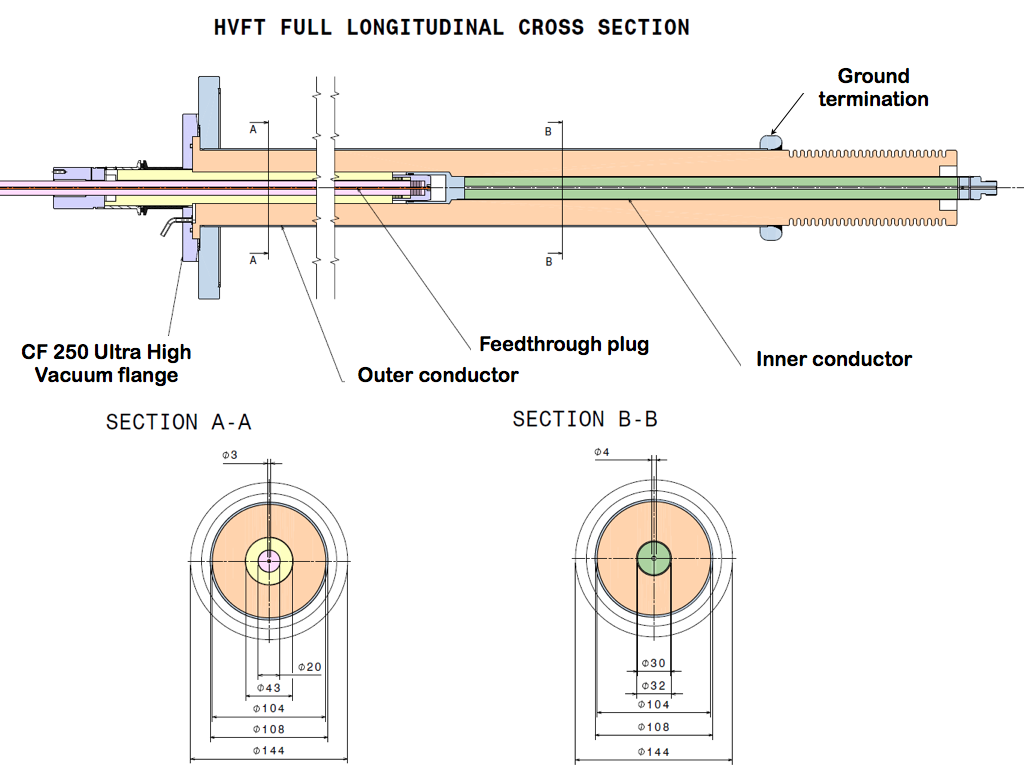}
  \caption{300 kV high voltage feedthrough cross-section.}
  \label{fig:HVFT-2D}
\end{figure}

A cross-section drawing of the 300 kV high voltage feedthrough is shown in \figref{fig:HVFT-2D}. 
Its general design comes from the experience acquired during the ICARUS experiment \cite{Amerio:2004ze} where lower high voltage feedthroughs were successfully tested up to -150 kV and operated for several years at -75~kV. The 300 kV feedthrough is a scaled up and improved design. The scale up was essentially based on the following considerations: 1. increase the length of the feedthrough to penetrate the larger thicknesses of the passively insulated membrane cryostats, 2. enlarge the diameter of the inside hole to accept cable plugs transporting higher voltages, 3. keep the heat input to a minimal value to avoid formation of gas argon bubbles in the vicinity of the high voltage, 4. preserve the vacuum tightness of the cryostat and 5. reduce as much as possible the electric field at the ground termination. The main differences between the two feedthroughs are summarized in \tabref{tab:CompHVFT}. 

\begin{table}
\begin{center}
\begin{tabular}{p{1.9cm}p{2cm}p{3cm}p{3cm}p{2cm}}
\hline
\hline
\bf{Feedthrough} & \bf{Length of HMDPE rod} & \bf{Inner conductor diameter} & \bf{Outer conductor diameter} &
\bf{Shape of ground termination} \\
\hline
\hline
ICARUS & 1.3 m & 12 mm & 64 mm & circular\\
This paper & 2 m & 32 mm & 108 mm & elliptical\\
\hline
\hline
\end{tabular}
\caption{Differences between ICARUS and the feedthrough described in this paper.}
\label{tab:CompHVFT} 
\end{center} 
\end{table}

\begin{figure}[htb]
      \centering
        \includegraphics[width=0.95\textwidth, trim={0cm 0cm 0cm 0cm}, clip=true]{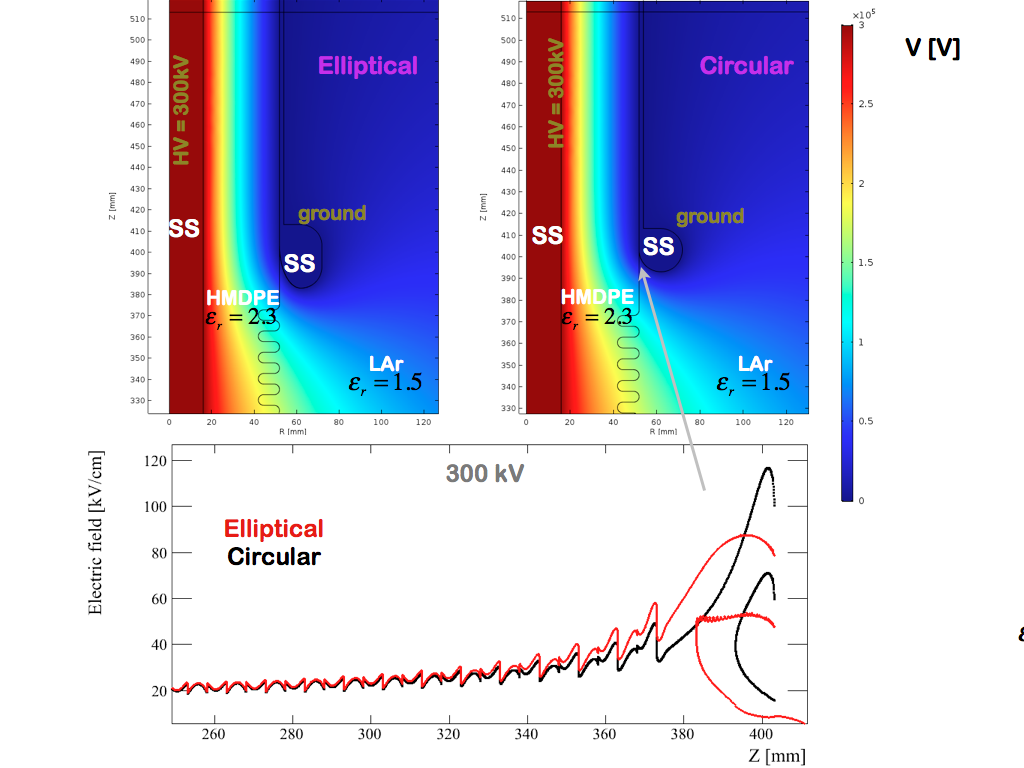}
      \caption{Electrostatic simulations in cylindric coordinates of the feedthrough in the region near the ground termination. The coloured surface represents the electric potential. {\it Left:} elliptical termination.  {\it Right:} circular termination. {\it Bottom:}
        electric field along the feedthrough as a function of z near the ground termination computed for an elliptical and a circular shape.}
      \label{fig:HVFTscheme}
  \end{figure}

The design follows a coaxial configuration with inner and outer conductors. The outer conductor consists of a 2 mm thick tubular shield with a CF250 Ultra High Vacuum flange welded on one end and an elliptical edge ring welded on the other.
The inner conductor consists of a 1 mm thick stainless-steel tube filled with FR4 to provide the vacuum tightness. It has a female receptacle to accept the feedthrough cable plug on one end and a male threaded contact on the other for electrical connection to the detector. Both inner and outer conductor tube thicknesses are chosen to minimise the heat input. The insulation between them is provided by one continuous  2 meter rod of High Molecular Density Polyethylene (HMDPE). We use HMDPE RCH-1000, which is rated for operations at -269$^{\circ}$ C, it has a relative dielectric permittivity \Er of 2.3 and a dielectric rigidity of 900 kV/cm. A very safe insulation strength of at least 2.7 MV is thereby provided by the 3 cm of HMDPE that separate the inner from the outer conductor. The area where the outer conductor is terminated though inevitably produces high electric fields. Electrostatic simulations show that the highest electric field in liquid argon is reached precisely in this particular area. As such great care should be taken to optimise the shape of the ground termination. For the ICARUS feedthrough the outer conductor ended with a ring of circular cross-section, whereas in the design of our feedthrough we adopted an elliptical ring based on results from electrostatic simulations. \figref{fig:HVFTscheme} shows the electric potential and the field with the inner conductor at 300 kV and the outer at ground. The simulations were performed with COMSOL\cite{comsolRef}. In the top plots we compare circular (right) and elliptical (left) ground terminations, where the coloured surface represents the value of the potential. The electric field along the side of the feedthrough as a function of the vertical coordinate $z$ near the ground termination is shown as well. It can be seen that the elliptical termination allows to reduce the deviation of the equipotential surfaces, thereby lowering the critical field by 30$\%$ from roughly 120 kV/cm to 80~kV/cm. 

\subsubsection{Manufacturing}

The first feedthrough was manufactured according to these design requirements by the company CINEL Strumenti Scientifici\footnote{\url{www.cinel.com/}}.
The first and most delicate step is to drill with high precision a
continuous hole along the 2 meters rod of HMDPE. The diameter of
the hole reduces from 43 to 32 mm to accept the feedthrough cable plug
on one end and to allow the insertion of the inner conductor on the
other. This step is quite critical mechanically, HMDPE is a rather
soft material and it is not trivial to drill a precise and aligned
hole in a 2 meter long rod. The inner conductor is then introduced in the rod and the last step consists in ''cryo-fitting'' them in the
outer conductor.

\begin{figure}[htb]
      \centering
      \includegraphics[width=1\textwidth,clip=true]{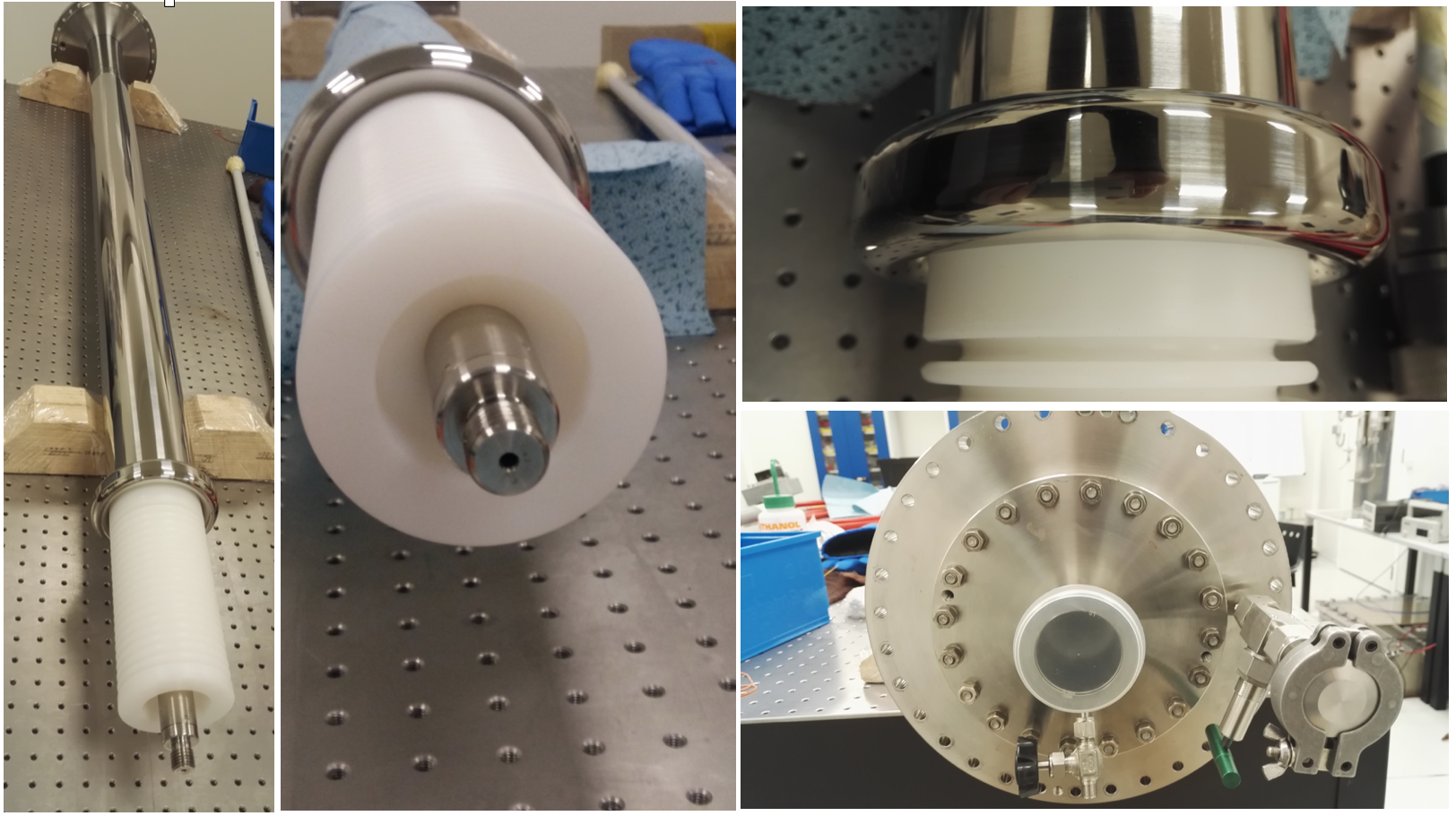}
      \caption{Pictures of the 300 kV high voltage feedthrough. {\it Left:} overall picture. {\it Centre:} bottom-end of the feedthrough. {\it Top-right:} elliptical shape ground ring and corrugated HMDPE insulator rod. {\it Bottom-right:} top part of the feedthrough with the CF-250 Ultra High Vacuum flange.}
      \label{fig:HVFTdrawing}
  \end{figure}

Pictures of the feedthrough are provided in \figref{fig:HVFTdrawing}.

The feedthrough vacuum tightness was measured by the company before delivery and again by ourselves. In both cases a Helium leak rate of less than 2.5 $\times$ 10$^{-10}$ mbar l/s was measured.

\subsection{Description of the high voltage power supply and cable}

\begin{figure}[htb]
  \centering
  \includegraphics[]{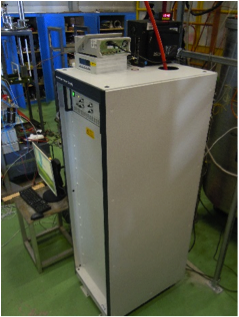}
  \includegraphics[]{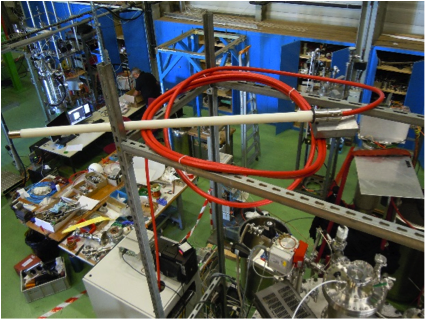}
  \caption{The Heinzinger 300 kV power supply during the test with the high voltage cable connected through the top.}
  \label{fig:HVpsu}
\end{figure}

The high voltage power supply used in this setup is the model PNChp 300000-05-neg from the company Heinzinger\footnote{\url{http://www.heinzinger.com}}. It is designed to supply -300 kV with a maximum current of 0.5 mA and a voltage residual ripple $\leq$ 0.001\% V$_{nom}$ $\pm$ 50 mV. It has an RS232 port as well as an analogue 0-10~V output for communications. The resolution is 1 $\mu$A and 1 kV in current and voltage respectively. The power supply is current limited. Once the current exceeds a set limit, the voltage drops until the current decreases below the set value. A picture is shown in \figref{fig:HVpsu}.

The high voltage cable follows a standard coaxial design with an external diameter of 22 mm and is also custom made by Heinzinger. The one
used in this setup has a length of 12 m, it is terminated on one end
by the male connection for insertion inside the power supply unit, and on the other end with a custom made plug designed to be inserted in the feedthrough.  
The plug that is inserted inside the feedthrough consists of a HMDPE insulated tube. When entering the plug the coaxial cable is stripped of its copper shielding which is firmly clamped on the stainless steel cap. The plug terminates with a stainless steel profile which is in electrical contact with the inner conductor of the cable. \tabref{tab:HVcable} summarises the technical specifications of the high voltage cable and the feedthrough plug.
\begin{table}[htb]
\begin{center}
\begin{tabular}{l|c} 
\hline
\hline
Rated voltage & 300 kV\\
Capacitance & 101 pF/m\\
Inductance & 0.3 $\mu$H/m\\
Center Core & Copper \\
Dielectric insulator  & Polyethylene\\
Woven copper shield & CuSn\\
Outer plastic jacket  & PVC \\
Colour & red\\
Minimum bending radius & 440 mm\\
Temperature resistance up to & ~60$^{\circ}$C\\
\hline
\hline
\end{tabular}
\caption{Specifications of the high voltage cable.}
\label{tab:HVcable} 
\end{center} 
\end{table}

\section{Description of the high voltage test} \label{sec:HVtest}
 \subsection{Experimental setup}\label{sec:ExperimentalSetUp}

\begin{figure}[htb]
     \centering
     \includegraphics[width=1\textwidth,clip=true]{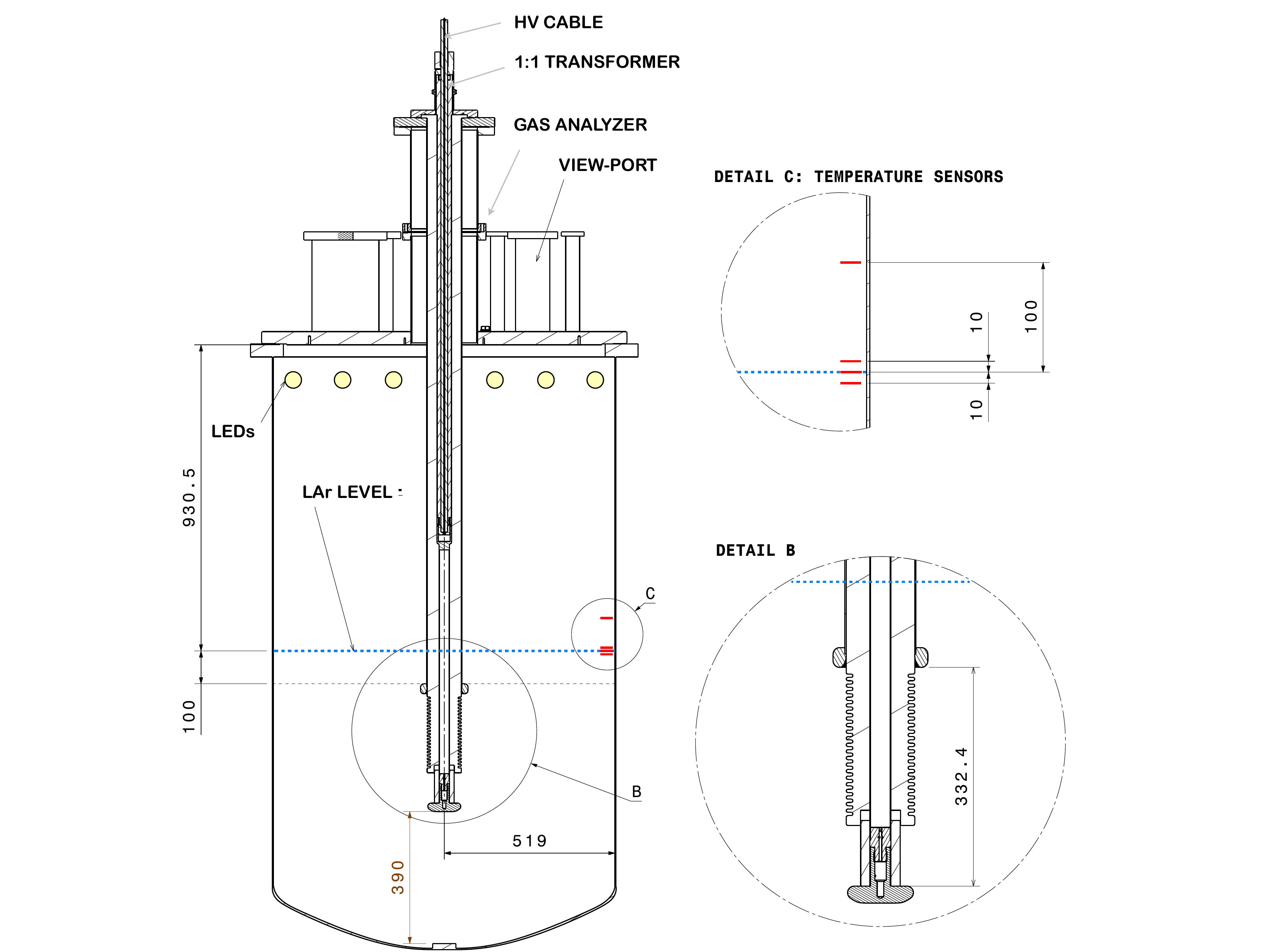}
     \caption{Schematic representation of the test setup.}
     \label{fig:SetUp}
   \end{figure}

The schematic representation of the test setup is illustrated in
\figref{fig:SetUp}. The setup consists of a 1 meter diameter vacuum insulated dewar with a 4 cm thick stainless steel cover that hosts the high voltage feedthrough through its center. Inside the dewar the feedthrough is terminated by a circular 10 cm$^2$ Rogowski shaped electrode \cite{RogowskiRef}. 
Some LEDs are installed in the upper part of the dewar. They can be switched on to help visual inspections of the LAr level and conditions inside the dewar. 
$O_{2}$ impurities are also monitored through a gas analyzer model AMI 2001RS, with a sensitivity of 100 ppb.

 Before the filling, the vessel is evacuable in order to remove air traces, favour the outgassing of the materials and check the absence of leaks to atmosphere. The dewar is filled with liquid Argon purified through a molecular sieve (ZEOCHEM Z3-06), which filters water molecules, and a custom-made copper cartridge, which removes oxygen and other electronegative molecules. 
The liquid Argon level is visually checked through a vacuum sealed viewport and estimated also using a temperature sensor placed in the vicinity of the requested LAr level. The nominal value of liquid Argon level used in the test is about 500 mm from the bottom of the dewar. In this configuration the liquid is 100 mm above the top of the termination of the ground outer conductor of the feedthrough (see \figref{fig:SetUp}). There is no active cooling of the liquid Argon. Instead the pressure inside the dewar is controlled by exhausting the boil off Argon through an external liquid argon bubbler. The  dominant heat input to the liquid argon comes from the non-insulated top cover situated at about 1~meter above the nominal level. 

The voltage is provided by the 300 kV power supply through a HV coaxial
cable. A 1:1 transformer coupled with the inner conductor of the cable
and insulated by the polyethylene shield of the cable has been added
to the cable to inductively couple the high voltage wire to an
oscilloscope. When a pulsed current flows through the cable, it is detected on a scope. This sensor is used to monitor the
frequency of possible micro discharges as well as the current delivered by the power supply during charging up. The residual heat input through conduction from the feedthrough in this particular setup was calculated to be $\simeq 1.5$~W.
 
\begin{figure}[htb]
     \includegraphics[width=1\textwidth, trim={0cm 1cm 0cm 0cm}, clip=true]{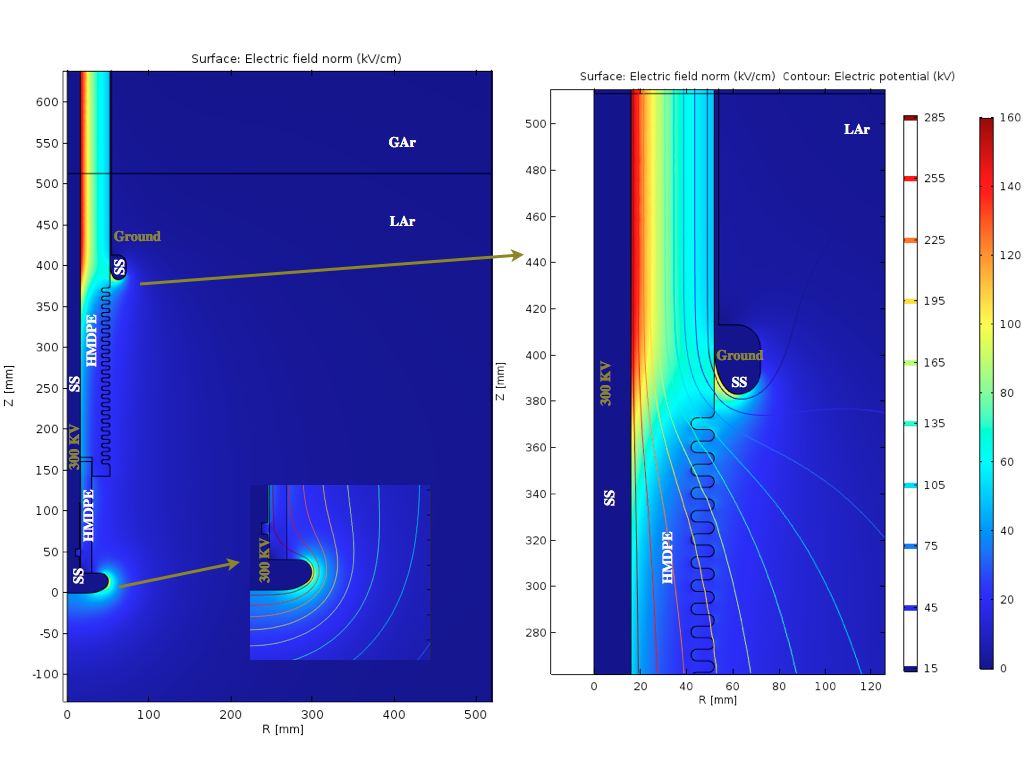}
     \caption{Electrostatic simulations of the high voltage test setup in cylindrical coordinates. The coloured surface corresponds to the electric field
       and the lines to the equipotentials.}
     \label{fig:2DSIM}
   \end{figure}

\begin{figure}[htb]
     \includegraphics[width=1\textwidth, trim={0cm 4.5cm 0cm 2cm}, clip=true]{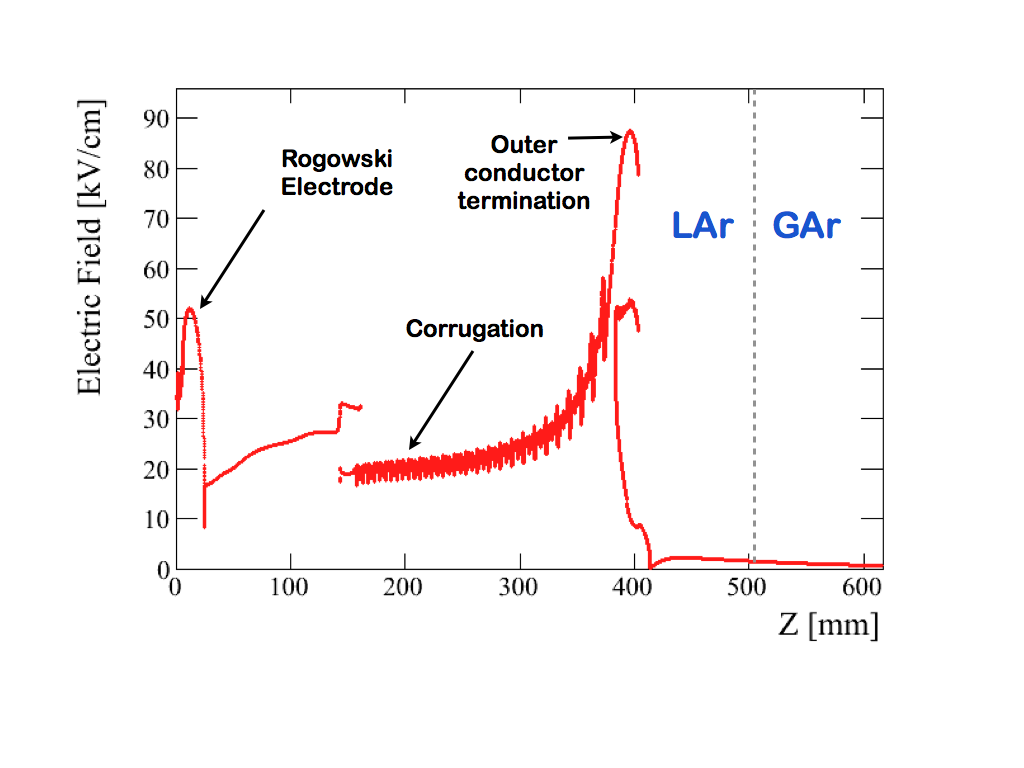}
     \caption{Electric field along the surface of the high voltage feedthrough.}
     \label{fig:fieldHVFT}
   \end{figure}

Electrostatic simulations of the full setup were performed to investigate regions of potentially high electric fields. The computation is presented in \figref{fig:2DSIM} considering a static voltage of 300 kV through the inner conductor with the cryostat walls and outer conductor at ground. The coloured surface represents the absolute value of the electric field. The higher field regions are localised near the termination of the Rogowski electrode and the termination of the feedthrough outer conductor, as is illustrated in the zoomed plots of \figref{fig:2DSIM}. The electric field following a line on the outer surface of the feedthrough is plotted in \figref{fig:fieldHVFT}. It can be seen that the highest surface field are reached on the ground elliptical termination of the outer conductor with a value of about 85~kV/cm and at the edge of the Rogowski profile where the field reaches up to 50~kV/cm.

\subsection{Results}

Two series of tests were performed in September 2016. The procedure consisted in ramping up the high voltage while constantly monitoring the values of the current and the high voltage to verify the absence of discharges. Large discharges would result in a significant increase of current and a voltage drop directly visible on the power supply display. The transformer installed in the high voltage cable allows for further monitoring of micro-discharges that would potentially be out of the resolution range of the power supply ($<1\mu$A). 

\begin{figure}[htb]
     \includegraphics[width=1\textwidth, trim={0cm 2.5cm 0cm 0cm}, clip=true]{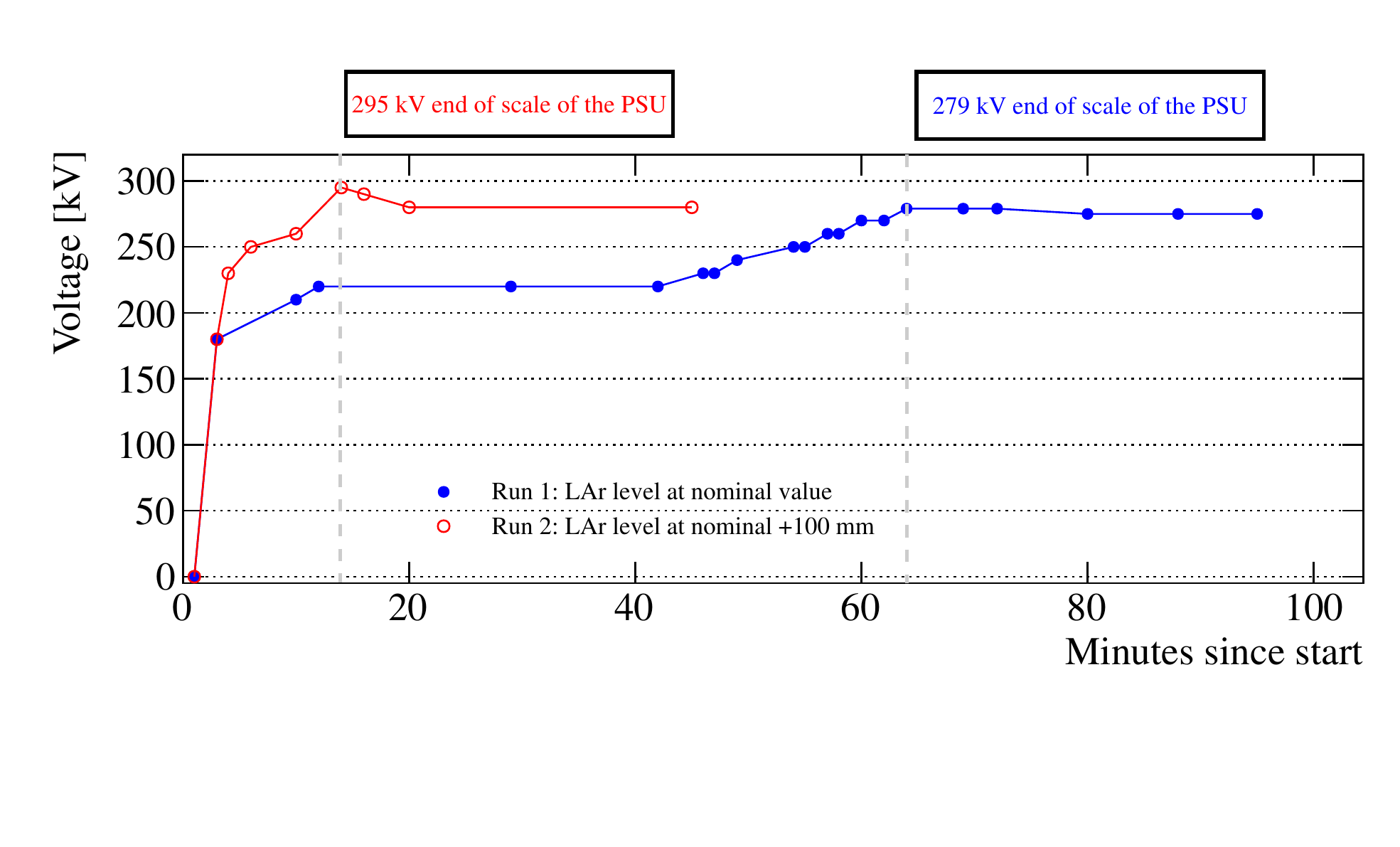}
     \caption{Applied High voltage as a function of time during the first two runs. The pressure was stable at 1.04 bar and O$_{2}$ impurities
       measured in the gas were below 100 ppb.}
     \label{fig:VoltageVsTime}
   \end{figure}

For the first series of tests we slowly raised the high voltage on the power supply at a rate of about -10 kV/min. Thanks to the bubbler, we were able to maintain the pressure inside the dewar 50~mbar above the atmospheric pressure. O$_{2}$ impurities were repeatedly measured in the Argon vapour and were found to be below 100 ppb. The surface of the liquid argon could be observed from the viewport and the absence of large waves or boiling could be checked visually. From time to time, a bubble would form around the feedthrough and would move upward along the side of the corrugated region, reaching the elliptical ground.
This exercise which was repeated twice is illustrated in \figref{fig:VoltageVsTime}. On both occasions we were able to reach the maximum voltage delivered by the power supply. 

A second series of tests were aimed at understanding the stability of the feedthrough during longer term operations. During these runs, we were able to ramp up until the end of the scale of the power supply in seconds. Table \ref{tab:HVLTtests} summarizes the voltages applied and the conditions during the three runs. In all cases we could reach and keep the voltage stable. 

At the highest settings of -250 and -275 kV we could observe sudden increases of the power supply current from 0 up to 2 $\mu$A, generally lasting less than a second before disappearing again. We called these fluctuations micro-discharges. Such micro-discharges occurred at random intervals and did not lead to large breakdowns.
We interpret them as transient currents flowing from the live Rogowski profile to the ground elliptical termination of the feedthrough. We could not discriminate whether these were surface currents along the corrugated HMDPE over a distance of 40 cm or directly through the liquid Argon. We note however that after the test, visual inspection of the corrugated region of the high voltage feedthrough did not show traces of damage or carbonisation. The Rogowski was also found to be unaltered. On the other hand we know that the thermodynamical conditions of the liquid argon in the test were not completely stable due to the lack of active cooling. As already mentioned in \secref{sec:introduction} even small variations of the LAr thermodynamical conditions can be a source of electrical instabilities, in particular for their tendency to trigger nucleation of bubbles in high electric field regions.
This potential issue should not be of concern for actual applications to LAr-TPC where the cathode will be located several meters away from the ground termination. We therefore conclude that the tests reported above demonstrate the feasibility of a feedthrough satisfying all our requirements up to 300 kV.

\begin{table}[htb]
\begin{center}
\begin{tabular}{|p{1.5cm}p{1.5cm}p{3cm}p{5cm}|}
\hline
\hline
\textbf{Voltage [kV]} & \textbf{Pressure [bar]} & \textbf{O$_{2}$ impurities in gas [ppm]} &
\textbf{Electric field at ground termination [kV/cm]}\\
\hline
-275 & 1.02 & $<0.1$ & 80\\
-250 & 1.06 & $<0.1$ & 73\\
-100 & 1.06 & $<0.1$ & 29\\
\hline
\hline
\end{tabular}
\caption{Long-term tests performed during one hour and with the liquid level 10 cm above the nominal value. The simulated electric field at the critical point (i.e. ground termination) it also shown for each settings (see text for more explanations).}
\label{tab:HVLTtests}
\end{center} 
\end{table}

\section{Conclusion}\label{sec:conclusions}

In this paper, we reported on the operation of a feedthrough up to -300 kV, which  represents a first milestone towards liquid Argon TPCs with very long drifts.
A high voltage feedthrough was designed and manufactured and shown to satisfy our requirements on length and diameter, vacuum tightness, thermal input and electrical specifications. Very high voltages could be applied during long periods repeatedly. A source of instability was observed which we called micro-discharges. We believe that these were specific to the setup configuration which was used for our test. They indicate that the distance between the live electrode and the ground termination of the feedthrough will have to be carefully defined in the future.

\section*{Acknowledgements}
This work would not have been possible without the support of the Swiss National Science Foundation and 
the ETH Zurich. One of us (L.M.B) is supported by the EC Horizon AIDA-2020 program. We also acknowledge the help of CERN
and its Neutrino Platform, where the tests were conducted.

\bibliographystyle{naturemag}
\bibliography{biblio_HVPaper}

\end{document}